\documentstyle[11pt]{article}
\begin{document}


\title{ 
\vspace*{-1cm}
\begin{flushright}
UR-1489
\end{flushright}
{\bf  ``Induced" Super-Symmetry Breaking
with a Vanishing Vacuum Energy}}
\author{Ashok Das and Sergio A. Pernice \\
\\
Department of Physics and Astronomy, \\
University of Rochester,\\
Rochester, New York, 14627. }
\maketitle

\begin{abstract}
A new mechanism for symmetry breaking is proposed which naturally
avoids the constraints following from the usual theorems of symmetry
breaking. In the context of super-symmetry, for example, the breaking
may be consistent with a vanishing vacuum energy. A 2+1
dimensional super-symmetric gauge field theory is explicitly shown to
break super-symmetry through this mechanism while maintaining a zero
vacuum energy. This mechanism may provide a solution to two long
standing problems, namely, dynamical super-symmetry breaking and the
cosmological constant problem.
\end{abstract}

\section{Introduction}

Dynamical supersymmetry breaking remains a fundamental problem
in particle physics~\cite{BW, PR}. As pointed out long ago~\cite{Z},
supersymmetry algebra implies that if supersymmetry
is unbroken, the vacuum must have a vanishing energy leading to
an elegant ``solution" of the cosmological constant problem~\cite{we}.
But if supersymmetry is a symmetry of the physical theories, it must
be broken, as is clear from the observed spectrum of particles. This,
again from the supersymmetry algebra, would imply that the vacuum
energy must be strictly positive - leading back to the cosmological
constant problem. This is one of many situations where it would
be desirable to break a symmetry bypassing the usual theorems on
symmetry breaking. It is, therefore, worth investigating alternate
scenarios where this might be achieved.

Let us recall some of the essential features of symmetry
breaking~\cite{golds, we2}. When
a symmetry is broken, there is a set of degenerate vacuum states
$| 0_n >$. For simplicity, let us consider the breaking of a discrete
symmetry with a finite number of such states which are assumed to be
orthonormal
\begin{equation}\label{orth}
< 0_n | 0_m > = \delta_{n,m}
\end{equation}
The symmetry operation mixes these states and, therefore, they are
degenerate. Every Hermitian local operator $A (x)$ of the theory will
have only diagonal matrix elements
\begin{equation}\label{hlo}
< 0_n | A (x) | 0_m > = \delta_{n,m} a_m
\end{equation}
The Hilbert spaces built on each of them are unitarily inequivalent.
The distinct Hilbert spaces are, therefore, dynamically disconnected.

We see, then, that symmetry breaking has two essential ingredients:
the existence of a symmetry of the theory, and the
splitting of the Hilbert space into dynamically separated sectors.
These two ingredients, however, need not always come together. The
Hilbert space of a theory may split into dynamically disconnected
parts {\it not} associated with an underlying symmetry (this, in
the language of statistical mechanics, just amounts to loss of
ergodicity, which is not necessarily associated with an underlying
symmetry). The usual theorems on symmetry breaking~\cite{golds}, assume
that the dynamically disconnected Hilbert spaces are mapped into
each other by the symmetry operation.
This is what actually happens in many field theories but it need not
always be the case. For example, suppose our theory has {\it two}
symmetries realized in the {\it full} Hilbert space and one is broken by the
usual mechanisms. In this case, the states of the Hilbert space built
on one of the degenerate vacua will carry representations of the broken
(residual) symmetry and, unless there are specific relations between the two
symmetries, may not contain all the states corresponding to the
representations of the second symmetry. In the process, the multiplets
of the second symmetry may get distributed into the dynamically
disconnected Hilbert spaces. This would, of course, ``induce" a
breaking of the second symmetry in the physical sector of the Hilbert
space and for such a breaking the usual theorems will no longer apply.
Thus, for example, if because of spontaneous breaking of some symmetry
in a theory, some of the superpartner states are removed from the
physical Hilbert space, supersymmetry will be broken. This, however,
would no longer necessarily imply positivity of the vacuum energy as
we will see. Such a scenario would have the added advantage that some
superpartner states would be completely unobservable (they do not belong
to the physical Hilbert space) consistent with experimental observations.
Furthermore, if such a phenomenon were to occur at low energies, the
beautiful short distance properties following from supersymmetry would
remain intact.

\section{Dynamical Supersymmetry Breaking}

In this letter we would like to explore the possible applications of these
general notions. In particular, we will study a supersymmetric field theory
where supersymmetry is explicitly shown to be broken while the vacuum
energy remains zero. Consider $2+1$ dimensional $N=2$ super-symmetric
Abelian Chern-Simons theory given by the action
\begin{eqnarray}\label{sscs}
S &=& \int  {\rm d}^3 x  \left\{ {\kappa \over 4 c} \ \varepsilon^{\mu \nu \lambda}
A_{\mu} F_{\nu \lambda} + |D_{\mu} \phi|^2 + i \ \overline{\psi} \ \gamma^{\mu}
D_\mu \psi  \right.  \nonumber \\
 & &  \qquad \ \left.  \rule{0in}{3.0ex}
- ( {e^2 \over \kappa c^2} )^2    | \phi |^2 \left[  | \phi |^2 - v^2  \right]^2
+  {e^2 \over \kappa c^2}  \left[  3 | \phi |^2 - v^2  \right]  \ \overline{\psi}
\psi  \right\}
\end{eqnarray}
where $D_{\mu} = \partial_{\mu} + i (e / c) A_\mu$, $\phi$ is a complex spin-0
field, $\psi$ is a complex two-component Dirac field corresponding
to spin $1/2$ particles, and $\kappa$ may be positive or negative. We
consider the positive case. There are therefore two degrees of freedom
corresponding to the complex scalar field and two degrees of freedom
corresponding to the fermion and its anti-particle (in 2+1 dimensions both
the fermion and its anti-particle have only one spin degree of freedom). The
Chern-Simons field $A_\mu$ carries no dynamical degrees of freedom in the
symmetric phase so the super-symmetric counting of degrees of freedom follows.
The mass of the supersymmetric partner particles is $m = e^2 v^2 / c^2 \kappa$.
The model of eq.~(\ref{sscs}) was first studied by Lee, Lee and Weinberg~\cite{llw}.

In Ref.~\cite{llm},  Leblanc, Lozano and Min studied the symmetries of  the
low energy effective theory corresponding to~(\ref{sscs}). We review some of
their results relevant for our work. In the low energy limit,~(\ref{sscs})
corresponds to a Galilean invariant $N=2$ supersymmetric field theory. In the zero
antiparticle sector (in the nonrelativistic limit particle and antiparticle numbers
are separately conserved) its Hamiltonian is given by
\begin{equation}\label{nrham}
H = \int {\rm d}^2 r  \ \left\{ {1 \over 2 m}\left[  \left( {\bf D} \Phi \right)^\dagger
\cdot {\bf D} \Phi + \left( {\bf D} \Psi \right)^\dagger {\bf D} \Psi \right] -
{e \over 2 m c}  : B \ \rho_F :  - {e^2 \over 2 m c \kappa} : \rho^2_B : - 3
{e^2 \over 2 m c \kappa} \rho_b \rho_F \right\}
\end{equation}
where
\begin{equation}\label{rho}
\rho_B = |\Phi|^2  \quad  {\rm and} \quad  \rho_F = |\Psi|^2 
\end{equation}
and the (scalar) magnetic field is given by $B = \nabla  \times {\bf A}$\footnote{
In two dimensions, the cross product of two vectors ${\bf V} \times {\bf W} =
\varepsilon^{ij} V_i W_j$ is a scalar,  the curl of a vector (also a scalar) is
$\nabla \times {\bf V} = \varepsilon^{ij} \partial_i V^j$, and the curl of a
scalar is a vector with components $( \nabla \times S)^i = \varepsilon^{ij}
\partial_j S$. }. The bosonic field $\Phi$ corresponds to the nonrelativistic
limit, in the zero antiparticle sector, of the original field $\phi$.The fermionic
field $\Psi$ corresponds also to the nonrelativistic limit, in the zero antiparticle
sector, of the first component of the of the two component original fermionic
field $\psi$. The second component has been eliminated through its equation
of motion. In the covariant derivatives the gauge field is replaced through its equation of
motion in Coulomb gauge~\cite{llm, ha1, jp} by
\begin{equation}\label{gfrepl}
{\bf A} (t, {\bf r}) = {e \over \kappa} \nabla \times \int {\rm d}^2 r'
G ({\bf r}' - {\bf r}) \ \rho (t, {\bf r}')
\end{equation}
where $\rho = \rho_B + \rho_F$ and $G ({\bf r})$ is the two dimensional 
Green's function of the Laplacian
\begin{equation}\label{lapl}
G ({\bf r}) = {1 \over 2 \pi} \ln{ | {\bf r} |}
\end{equation} 

The model defined by the Hamiltonian~(\ref{nrham}) corresponds to the
super-symmetric extension of  a model studied by Jackiw and Pi~\cite{jp}.
We see then from~(\ref{nrham}) that the nonrelativistic limit of~(\ref{sscs})
corresponds to bosons interacting (minimally) with the gauge field and with
themselves through a contact interaction. The fermions also interact minimally
with the gauge fields but in addition they posses a Pauli interaction. This
non-minimal  interaction arose algebraically from the elimination of the second component of the spinor field through its equation of motion, but its
appearance can also be expected, on general grounds, in the nonrelativistic
limit of super-symmetric gauge theories~\cite{dp}. Finally, there is a contact
boson-fermion interaction.

In~\cite{llm} it was shown that the model~(\ref{nrham}) has 16 generators
of  symmetry operations that generate an extended super-conformal
Galilean algebra. Of immediate interest to us is the algebra of the generators
of supersymmetry. It corresponds to a $N=2$ Galilean supersymmetry. The two
supercharges in the nonrelativistic theory are given respectively by\footnote{
For any vector ${\bf V} = (V^1 , V^2)$, $V^\pm = V^1 \pm i V^2$.}
\begin{eqnarray}\label{nrsusyQ1}
Q_1 &=& i \sqrt{2 m} \int {\rm d}^2 r  \ \Phi^\dagger \Psi  \\  \label{nrsusyQ2}
Q_2 &=& i {1 \over \sqrt{2 m}} \int {\rm d}^2 r  \ \Phi^\dagger D_+ \Psi 
\end{eqnarray}
and together with the Hamiltonian~(\ref{nrham}) they satisfy the super
symmetry algebra
\begin{eqnarray}\label{nrsusyalg1}
\left\{ Q_1, Q_1^\dagger \right\} &=& 2  M  \\  \label{nrsusyalg2}
\left\{ Q_2, Q_2^\dagger \right\} &=& H  \\   \label{nrsusyalg3}
\left\{ Q_1, Q_2^\dagger \right\} &=& P_-  \\   \label{nrsusyalg4}
\left\{ Q_\alpha , Q_\beta \right\}  &=&
\left\{ Q_\alpha^\dagger ,  Q_\beta^\dagger \right\}  = 0
\end{eqnarray}
The mass operator in eq.~(\ref{nrsusyalg1}), which commutes with all other
generators, is necessary in a Galilean invariant theory to close the
algebra~\cite{bar}.

According to the Bargmann superselection rule~\cite{bar},
a nonrelativistic supersymmetric field theory should be equivalent to a
supersymmetric Schr\"{o}dinger  equation in each particle number sector
of the theory. We will analyze in particular the two particle sector, that is both
tractable and non-trivial, which can be derived from the above expressions
assuming the existence of a zero particle, zero energy vacuum. Following
ref.~\cite{llm}, let us define the orthonormal two particle states $| E , N_B ,
N_F >$ with energy $E$, $N_B$ number of bosons and $N_F$ number of
fermions as follows
\begin{eqnarray}\label{2pst}
| E , 2, 0 >  & = &  {1 \over \sqrt{2}} \int  {\rm d} r_1  {\rm d} r_2 \
u_B ({\bf r_1}, {\bf r_2})  \Phi^\dagger ({\bf r_1})  \Phi^\dagger ({\bf r_2})
| \Omega >    \\
| E , 0, 2 >  & = &  {1 \over \sqrt{2}} \int  {\rm d} r_1  {\rm d} r_2 \
u_F ({\bf r_1}, {\bf r_2})  \Psi^\dagger ({\bf r_1})  \Psi^\dagger ({\bf r_2})
| \Omega >    \\
| E , 1, 1 >  & = & {1 \over \sqrt{2}} | E , 1, 1 >_S + {1 \over \sqrt{2}} 
| E , 1, 1 >_A  \nonumber  \\
& = & {1 \over 2} \int  {\rm d} r_1  {\rm d} r_2 \
\left\{  u_S ({\bf r_1}, {\bf r_2}) \left[  \Phi^\dagger ({\bf r_1})  \Psi^\dagger
({\bf r_2}) + \Phi^\dagger ({\bf r_2})  \Psi^\dagger ({\bf r_1}) \right] | \Omega >
 \right.  \nonumber \\
& &  \qquad\qquad\quad \left. \rule{0in}{3.0ex} + u_A ({\bf r_1}, {\bf r_2}) 
\left[  \Phi^\dagger ({\bf r_1})  \Psi^\dagger ({\bf r_2}) - \Phi^\dagger 
({\bf r_2})  \Psi^\dagger ({\bf r_1}) \right] | \Omega >  \right\}
\end{eqnarray}
The one fermion, one boson state is then divided into its symmetric and its
antisymmetric part  under the interchange of the two particles.

In the basis
\begin{equation}\label{ssbasis}
\Upsilon =  \left (\matrix{u_B\cr u_S\cr u_A\cr u_F\cr }\right )
\end{equation}
the supercharges~(\ref{nrsusyQ1}) and~(\ref{nrsusyQ2}) have the form
\begin{eqnarray}\label{sch12ps}
Q_1 = 2 i \sqrt{m} \left (\matrix{
0 & 1 & 0 & 0 \cr
0 & 0 & 0 & 0 \cr
0 & 0 & 0 & 1 \cr
0 & 0 & 0 & 0 \cr
}\right )\quad  ,  \qquad   
Q_2 = {1 \over 2  \sqrt{m}} \left (\matrix{
0 &  {\cal D}_1^+ + {\cal D}_2^+ & - {\cal D}_1^+ + {\cal D}_2^+ & 0 \cr
0 & 0 & 0 & {\cal D}_1^+ - {\cal D}_2^+ \cr
0 & 0 & 0 & {\cal D}_1^+ + {\cal D}_2^+ \cr
0 & 0 & 0 & 0 \cr
}\right ) 
\end{eqnarray}
where
\begin{eqnarray}\label{covdearqm}
{\cal D}_1 = \nabla_1 - {i e^2  \over c \kappa} \nabla_1 \times
G ({\bf r_1} - {\bf r_2}) \quad , \qquad
{\cal D}_2 = \nabla_2 - {i e^2  \over c \kappa} \nabla_2 \times
G ({\bf r_2} - {\bf r_1})
\end{eqnarray}
The two particle Hamiltonian, in the  same basis, is
\begin{eqnarray}\label{schh}
H =  \left\{ - {1 \over 2 m}  \left(  {\cal D}_1^2 + {\cal D}_2^2 \right) I
- {e^2 \over  m c \kappa} \ \delta ({\bf r_1} - {\bf r_2})
\left (\matrix{
1 & 0 & 0 & 0 \cr
0 & 1 & 0 & 0 \cr
0 & 0 & -1 & 0 \cr
0 & 0 & 0 & -1 \cr
}\right ) \right\}
\end{eqnarray}
From eqs.~(\ref{sch12ps}) and~(\ref{schh}) one can easily check that
the algebra of eqs.~(\ref{nrsusyalg1}-\ref{nrsusyalg4}) is satisfied with
the mass operator corresponding to $M = 2 m I$ as it should be in the 
two particle sector.

From eq.~(\ref{sch12ps}) and~(\ref{covdearqm}) it is clear that the  $12$ and
the $34$ components of  $Q_2$ refer to the center of mass momentum and
have no dynamical effect.  In order to investigate the dynamical content
of the system under consideration we will, therefore, work in the center of mass
reference frame where they vanish. In that reference frame, in terms of the
relative variable ${\bf r} = {\bf r_1} - {\bf r_2} $ and in polar coordinates,
the only nonvanishing components of $Q_2$ have the form
\begin{equation}\label{q213}
Q_{2}^{13} = - Q_{2}^{24} = - {e^{i \theta} \over \sqrt{m}}   \left[ {\partial \over
\partial r} + { 1 \over r}  i {\partial \over \partial \theta}  - {e^2 \over c \kappa}
{\partial G \over \partial r} \right]
\end{equation}
Similarly,  for $Q_2^\dagger$ the only nonvanishing  components are
(remember that in two dimensions  $( \partial / \partial r )^\dagger  =  -\partial
/ \partial r - 1 / r$)
\begin{equation}\label{q213dagger}
\left( Q_{2}^\dagger \right)^{31} = - \left( Q_{2}^\dagger \right)^{42} = -
{1  \over \sqrt{m}}   \left[ - {\partial \over \partial r} +
{1 \over r} \left( i {\partial \over \partial \theta}  -1 \right) - {e^2 \over c \kappa}
{\partial G \over \partial r} \right] e^{- i \theta}
\end{equation}
From eqs.~(\ref{q213}) and~(\ref{q213dagger}), remembering
that these charges are defined in the basis given in eq.~(\ref{ssbasis}),
we find the following relations between superpartner states:
the superpartner of a boson-boson state with orbital angular momentum
$2 \ell$, $|bb, 2 \ell >$ (i.e., the state, $Q_{2}^\dagger \ |bb, 2 \ell >$),
is a
boson-fermion antisymmetric state with orbital angular momentum $2 \ell +1$,
that is, $|bf_A, 2 \ell+1>$. A similar relation holds for a  boson-fermion symmetric
state $|bf_S, 2 \ell>$ and its fermion-fermion superpartner.  The difference in
one unit of orbital angular momentum for superpartner states (which is
perfectly compatible with $\Delta J = 1/2$ in relativistic field theories)
was shown to be
a general property of two dimensional supersymmetric quantum mechanics
in ref.~\cite{dp}. This property holds for the supercharge related to the
Hamiltonian in the supersymmetry algebra  ($Q_2$ in our case). In terms of
the general formalism of ref.~\cite{dp}, the interaction part of the supercharges
corresponds to a vector superpotential $\vec{W} = i (e^2 / c \kappa)  
({\partial G /  \partial r}) \ \hat{\theta}$.  $\hat{\theta}$ refers to the unit vector
in the angular direction.

We will now study the eigenvalue equations for the superpartners. The 
analysis for the super pair boson-boson and boson-fermion antisymmetric
is identical to the other super pair (boson-fermion symmetric and
fermion-fermion), so we only analyze the former. The eigenvalue
equation for the state $u_B ({r}) \ e^{-  i 2 \ell \ \theta} = < {\bf r} | \ bb, 2 \ell>$
is
\begin{equation}\label{bb2meq}
Q_{2}^{13}  \left( Q_{2}^\dagger \right)^{31}  u_B ({r}) = 
{1 \over m} \left[  - {\partial^2 \over \partial r^2} - {1 \over r}
{\partial \over \partial r} + {\left( 2 \ell - \lambda \right)^2 \over r^2}
- \lambda { \delta (r) \over r} \right] u_B = E u_B ({r})
\end{equation}
where $\lambda = e^2 / (2 \pi c \kappa)$. On the other hand, the
superpartner state
\begin{equation}\label{uA}
u_A ({r}) \ e^{-  i (2 \ell + 1) \ \theta} = < {\bf r} | 
\  \left( Q_{2}^\dagger \right)^{31} | bb, 2 \ell> =   < {\bf r} | 
\ bf_A, 2 \ell + 1 >
\end{equation}
satisfies the equation
\begin{equation}\label{bf2meq}
\left( Q_{2}^\dagger \right)^{31}  Q_{2}^{13}  \ u_A ({r}) = 
{1 \over m} \left[  - {\partial^2 \over \partial r^2} - {1 \over r}
{\partial \over \partial r} + {\left( 2 \ell + 1 - \lambda \right)^2 \over r^2}
+ \lambda { \delta (r) \over r} \right] u_A = E u_A ({r})
\end{equation}
As we see, the boson-boson state feels an attractive delta function
interaction while the boson-fermion antisymmetric state a repulsive
one. In this second case, the delta function might be omitted since,
an antisymmetric $u_A$ would make it vanish at the origin. On the other
hand, $u_A$ is the superpartner of  $u_B$, so it must be generated
from it through the action of $Q_2^\dagger$. It is therefore more natural
to leave it in the equations. As we will see shortly, the sign of the delta
function interaction does not have any relevance. Whether attractive or
repulsive, the dynamical effect of it is to suppress non-regular solutions
(i.e., solutions not vanishing at the origin). Therefore everything is consistent,
and as we will see, a great surprise results from this consistency.
Equations~(\ref{bb2meq}) and~(\ref{bf2meq}) clearly show that, apart from
the delta function interactions, the dynamics of this model corresponds to a
supersymmetric Aharonov-Bohm problem~\cite{ahb}. This is not surprising since the (nonsupersymmetric)
Galilean invariant Chern-Simons theory is a field theoretic representation
of this problem~\cite{ha1}.

In the Aharonov-Bohm context, the solutions of  eqs.~(\ref{bb2meq})
and~(\ref{bf2meq}) are set to zero at the origin. This makes physical sense
because the solenoid where the magnetic field is confined is suppose to
be isolated from the particle by some other interaction (this interaction
forces the wave function to vanish at the origin). The purpose of Aharonov
and Bohm was precisely to show that in quantum mechanics the {\it potential}
has a physical effect, even when the particle is never in contact with the
magnetic field. In our context, however, equations~(\ref{bb2meq})
and~(\ref{bf2meq}) were derived as the nonrelativistic limit of  a
relativistic supersymmetric field theory. This field theory is suppose to
encode the totality of the interactions of the system represented by it.
Therefore, we cannot force by hand the behavior of the wave functions
at the origin - specially for the symmetric boson-boson case. However, as
mentioned before, the delta function interaction dictates {\it dynamically}
the behavior at the origin. It forces the solutions to vanish there, as is the case
(by hand) in the Aharonov-Bohm problem. For the boson-fermion
antisymmetric case, this is just a self consistent result. But for the
boson-boson case it involves and independent dynamical effect. So,
although the solutions of eqs.~(\ref{bb2meq}) and~(\ref{bf2meq}) are well
known, it is worth re-deriving them here once more in a form that 
explicitly exposes the aspects relevant for our purpose.

As mentioned, the sign and the magnitude of the coefficient of the
delta function turns out to be irrelevant (if different from zero). Therefore
let us consider the equation  
\begin{equation}\label{generic}
 \left[  - {\partial^2 \over \partial r^2} + { \alpha^2  - 1/4 \over r^2}
+ \beta { \delta (r -R) \over R} - E \right] v = 0
\end{equation}
that includes all possible cases with the identification $v = u   
\sqrt{r}$ and $\alpha = 2 \ell - \lambda$, $\beta = - \lambda$ for the
boson-boson case and $\alpha = 2 \ell + 1 - \lambda$, $\beta = \lambda$
for the boson-fermion antisymmetric case. The delta function has been
regularized; at the end of the calculations we will take the $R \rightarrow
0$ limit. The results should be independent of the particular
regularization chosen~\cite{ha2}.

Choosing, for $r \neq R$, a solution of the form 
\begin{equation}\label{sollprop}
v = \sum_{n=0}^{\infty} a_n r^{n+\delta}
\end{equation}
one finds $\delta = 1/2 \pm | \alpha |$. While for the plus sign the
corresponding solution $u = v / \sqrt{r}$ vanishes at the origin, for the
minus sign it diverges there.
However, note that if $| \alpha | < 1$  the minus sign solution is still
normalizable. Without making any assumption about the behavior at the
origin we have
\begin{eqnarray}\label{noassumpt1}
v (r) = A \sum_{n=0}^{\infty} a_{2n}^+ \  r^{2n + 1/2 +| \alpha |} +
B \sum_{n=0}^{\infty} a_{2n}^-  \ r^{2n +1/2 -|\alpha|} \quad , 
\qquad {\rm for} \quad  r > R   \\   \label{noassumpt2}
v (r) = C \sum_{n=0}^{\infty} a_{2n}^+ \  r^{2n +1/2 +| \alpha |} +
D \sum_{n=0}^{\infty} a_{2n}^-  \ r^{2n + 1/2 -|\alpha|} \quad , 
\qquad {\rm for} \quad  r < R 
\end{eqnarray}
where the coefficients $a_{2n}^+$ and $a_{2n}^-$ satisfy the relation
\begin{eqnarray}\label{coef1}
a_{2n + 2}^+ =  { - E \over  \left( 2n + 2 + 2 |\alpha| \right)
\left( 2n + 2 \right) } a^+_{2n}   \\  \label{coef2}
a_{2n + 2}^- =  { - E \over  \left( 2n + 2 - 2 |\alpha| \right)
\left( 2n + 2 \right) } a^-_{2n}
\end{eqnarray}
We see from these equations that the energy has to be positive or
zero. If it is negative the sign of the coefficients would not oscillate
and the wave functions would grow exponentially for large $r$.

At $r = R$, which we assume to be arbitrarily small (remember that the
limit $R \rightarrow 0$ is supposed to be taken at the end of the calculation),
the following two equations have to be satisfied
\begin{eqnarray}\label{constr1}
v (R+ \epsilon)  &=& v (R - \epsilon)  \\  \label{constr2}
v' (R - \epsilon) - v' (R + \epsilon) &=& {\beta \over R} v (R)
\end{eqnarray}
Let us, for simplicity of analysis, assume that $|\alpha| < 1$.The continuity
equation implies $(A-C) R^{2 |\alpha|} = (D-B)$. In the
limit $R \rightarrow 0$ this leads to $B=D$, with no constraints over
$A$ or $C$. Equation~(\ref{constr2}), together with the above result, implies
$\left[  (1/2 + |\alpha|) (C-A) + \beta C \right] R^{2 | \alpha |} + \beta D =0$.
In the limit $R \rightarrow 0$ it imposes $\beta D =0$. If $\beta \neq 0$,
independent of its sign, this implies $D=0$, and, therefore, $B=0$ as well.
We can easily show that this conclusion holds for any $|\alpha|$.

We see then that the {\it dynamics} determine the behavior at the origin.
For $|\alpha| > 1$, the irregular solutions could have been discarded
invoking normalizability. It is nice that the dynamics gets rid of such
solutions.
For $|\alpha| < 1$, normalizability alone is not enough to discard the
irregular solutions. Even more, the divergence of the wave function at
the origin could have been understood as the effect of the delta function
interaction in the attractive case (corresponding to the boson-boson states).
However, as we just saw,  whether attractive or repulsive, the effect of the
delta function interaction is to force the irregular solutions out. 
This is also consistent with the antisymmetric nature of the boson-fermion
states. So we conclude that in the general case, only solutions vanishing
at the origin survive. 

Remember that equation~(\ref{generic}) represents (in the reduced 
variables) equation~(\ref{bb2meq}) for a boson-boson state of
orbital angular momentum $2 \ell$ with the identification  $\alpha =
2 \ell - \lambda$. It also represents equation~(\ref{bf2meq}) for the boson-fermion antisymmetric superpartner state of  orbital angular
momentum $2 \ell +1$ with the identification $\alpha = 2 \ell +1 -
\lambda$. For our purposes it is convenient to consider
both regular and irregular solutions, keeping in mind that the irregular
ones do not belong to the physical Hilbert space. For definiteness,
let us assume
\begin{equation}\label{deff}
0 < \lambda = {e^2 \over 2 \pi c \kappa} < 1
\end{equation}
Consider the regular solutions corresponding to the boson-boson
state of orbital angular momentum $2 \ell$ for $\ell = 0, -1, -2, \cdots$.
Independent of their energy, at the origin, these solutions go to zero
as $r^{- \left( 2 \ell - \lambda \right)}$.
The superpartner solutions, that is, the solutions obtained by applying
$\left( Q_{2}^\dagger \right)^{31}$ to them, have orbital angular momentum
$2 \ell + 1$ and behave near the origin as $r^{- \left( 2 \ell + 1 - \lambda \right)}$.
For $\ell = 0$, again independent of their energy, this corresponds to an
irregular (but normalizable) solution that, as we have shown, does not
belong to the physical Hilbert space. Therefore, the boson-boson states
of {\it arbitrary energy} with zero orbital angular momentum do not have
a superpartner! Similarly, consider now the irregular solutions (that do not
belong to the physical Hilbert space) with arbitrary energy and negative or
zero orbital angular momentum. Near the origin they behave as $r^{ \left( 2 \ell -
\lambda \right)}$. Note that for $\ell = 0$, although not in the physical Hilbert
space, the solution is normalizable. Applying $\left( Q_{2}^\dagger \right)^{31}$
to them, we obtain solutions of the superpartner equation that near the
origin behave as $r^{ \left( 2 \ell +1 - \lambda \right)}$. For $\ell = 0$ the
exponent is positive, therefore, it corresponds to a regular solution that
belongs to the physical Hilbert space. We conclude then that the boson-fermion
antisymmetric states of orbital angular momentum equal to one and arbitrary
energy do not have a superpartner! Furthermore, the zero orbital angular
momentum boson-boson solutions and the boson-fermion antisymmetric
solutions of orbital angular momentum equal to one that do belong to the
physical Hilbert space are {\it not} superpartners of each other. The dynamical
separation of the physical Hilbert space produced by the delta function
interaction, among other things, pushes the irregular (but normalizable)
solutions outside the physical Hilbert space,  and induced the breaking of
supersymmetry!

Let us show explicitly the above statements.
From eqs.~(\ref{sollprop}),~(\ref{coef1}) and~(\ref{coef2}), remembering
that $\delta = 1/2 \pm |\alpha|$, that $u = v / \sqrt{r}$, and choosing
for convenience $a_0^\pm = \sqrt{E}^{\ \pm |\alpha|} / (\pm 2|\alpha|)$,
we obtain for the regular ($+$) and the irregular ($-$) solutions, the expressions
\begin{eqnarray}\label{solregirr}
u^\pm (r) = \sum_{n=0}^\infty  { (-1)^n \over (2 n)!! (2n \pm 2 |\alpha|)!!}
 \left( \sqrt{E} r \right)^{2n \pm |\alpha|}
\end{eqnarray}
where $(2n \pm 2 |\alpha|)!! = (2n \pm 2 |\alpha|) (2n-2 \pm 2 |\alpha|)
\cdots (\pm 2 |\alpha|)$ and similarly for $(2n)!!$. Therefore, for a
boson-boson state with zero or negative orbital angular momentum
(zero or negative $\ell$),
the regular and irregular solutions are, respectively,
\begin{eqnarray}\label{bbsolreg}
u_B^+ ({\bf r}) = \sum_{n=0}^\infty  { (-1)^n \over (2 n)!! (2n - 2 \left( 2 \ell -
\lambda \right))!!}  \left( \sqrt{E} r \right)^{2n -  \left( 2 \ell - \lambda
\right)}  e^{-i 2 \ell \theta} \\ \label{bbsolirr}
u_B^- ({\bf r}) = \sum_{n=0}^\infty  { (-1)^n \over (2 n)!! (2n + 2 \left( 2 \ell -
\lambda \right))!!} \left( \sqrt{E} r \right)^{2n +  \left( 2 \ell - \lambda
\right)} e^{-i 2 \ell \theta}
\end{eqnarray}
When we act on them with $\left( Q_{2}^\dagger \right)^{31}$, given in
equation~(\ref{q213dagger}), we obtain
\begin{eqnarray}\label{Q2bbsolreg}
\left( Q_{2}^\dagger \right)^{31} u_B^+ ({\bf r}) \propto \sum_{n=0}^\infty
{ (-1)^n  (2n - 2 \left( 2 \ell - \lambda \right)) \over (2 n)!! (2n - 2 \left( 2
\ell - \lambda \right))!!}  \left( \sqrt{E} r \right)^{2n - \left( 2 \ell - \lambda
\right) -1 }  e^{-i \left( 2 \ell +1 \right) \theta} \\ \label{Q2bbsolirr}
\left( Q_{2}^\dagger \right)^{31} u_B^- ({\bf r}) \propto \sum_{n=0}^\infty
{ (-1)^n \ 2n \over (2 n)!! (2n + 2 \left( 2 \ell - \lambda \right))!!} \left( \sqrt{E} r
\right)^{2n + \left( 2 \ell - \lambda \right) -1} e^{-i \left( 2 \ell +1 \right) \theta}
\end{eqnarray}
The regular solution then trivially can be written in the form
\begin{equation}\label{finalreg}
\left( Q_{2}^\dagger \right)^{31} u_B^+ ({\bf r}) \propto \sum_{n=0}^\infty
{ (-1)^n \over (2 n)!! (2n - 2 \left( 2 \ell +1 - \lambda \right))!!}  \left( \sqrt{E} r
 \right)^{2n - \left( 2 \ell +1  - \lambda \right) }  e^{-i \left( 2 \ell +1 \right) \theta} 
\end{equation}
which corresponds to the regular solutions of orbital angular momentum
$2 \ell + 1$ except for $\ell = 0$. In that case the solution becomes
irregular and does not belong to the Hilbert space as mentioned above.

For eq.~(\ref{Q2bbsolirr})  the situation is only slightly more involved.
Being the coefficients proportional to $2n$, the $n=0$ one vanishes
while in the others $2n / (2n)!! = 1/(2(n-1))!!$. Shifting variables with
$j = n-1$ and calling again $n$ to $j$ we obtain 
\begin{equation}\label{finalirr}
\left( Q_{2}^\dagger \right)^{31} u_B^- ({\bf r}) \propto \sum_{n=0}^\infty
{ (-1)^n \over (2 n)!! (2n + 2 \left( 2 \ell +1 - \lambda \right))!!} \left( \sqrt{E} r
\right)^{2n + \left( 2 \ell +1 - \lambda \right)} e^{-i \left( 2 \ell +1 \right) \theta}
\end{equation}
They correspond to irregular solutions with orbital angular momentum
equal to $2 \ell +1$ (remember that $\ell$ is negative or zero) except
for $\ell = 0$. In that case, eq.~(\ref{finalirr}) corresponds to a regular
solution and therefore belongs to the physical Hilbert space. This
finishes the proof of the breaking of supersymmetry.

\section{Discussion}

So, as we have seen, the whole tower of boson-boson states of zero orbital
angular momentum and arbitrary energy do not have superpartners. Similarly,
the entire tower of boson-fermion antisymmetric states of orbital angular
momentum equal to one and arbitrary energy do not have superpartners.
Clearly the same is true for the other supersymmetric pair as well. The
supersymmetry, $Q_2$, of the Hamiltonian is broken, but this does not
imply a Goldstino in the spectrum. Note that
in the context of a nonabelian gauge interaction, where we expect a
confining superpotential, the states without superpartners (corresponding
to lowest angular momentum) would naturally correspond to the lower
energy bound states. Therefore we would expect a spectrum consisting
of low energy bound states without superpartners. This is, of course,
what is to be expected on physical grounds. 

In order to derive the equations in the two particle sector,
the existence of a zero energy vacuum state was assumed. The situation in
the two particle sector, where the ground state is clearly a zero
(nonrelativistic) energy state, is illuminating in this respect. In this
sector we bypass the positivity of the ground state energy (that ``follows"
from the nonrelativistic supersymmetry algebra) under broken supersymmetry
because, as we have just seen, the breaking is of a very special nature.
Here, some of the superpartner states do not belong to the physical Hilbert
space. They are not in the domain  of the supercharge operators (a similar
observation was noted in connection with one dimensional supersymmetric
quantum mechanics in ref.~\cite{jr}). A vanishing ground state energy is
therefore perfectly consistent.

It is important to understand the underlying reasons for the ``induced"
supersymmetry breaking in order to look for realistic theories where
similar mechanisms might be operating. In this connection, let us note
that the model~(\ref{sscs}), is not invariant under parity.
Another way of saying the same is that a parity operation leads to
a different theory. From eqs.~(\ref{bbsolreg}-\ref{bbsolirr}) we note
that for $l=0$ (the supersymmetry breaking states), when $\lambda$
goes to $- \lambda$ (which is reminiscent of the parity transformation),
the regular solution
goes into the irregular one. So the induced breaking of supersymmetry is
very likely related to the parity violation in this theory. This issue,
however, deserves a much more careful and detailed investigation which we leave
for another paper. For the moment let us simply point out that in nature
parity is not conserved. It is suggestive and would, therefore, be
interesting if parity (or an alternate discrete symmetry such as CP)
violation can be tied to ``induced" supersymmetry breaking in
relativistic $3 + 1$ dimensional theories. This would not only explain
why superpartner particles have been so elusive but also would lead to
a solution of the cosmological constant problem. At present, we are trying
to find a realistic model which would incorporate ``induced"
supersymmetry breaking.

 \section*{Acknowledgments}

One of  us  (S.P.) would like to acknowledge useful discussions
with C. R. Hagen. This work was supported in part by the U.S. Dept. of
Energy Grant  DE-FG 02-91ER40685.

\end{document}